\documentclass[%
 reprint,
 amsmath,amssymb,
 aps,
prb,
]{revtex4-2}

\usepackage{graphicx}
\usepackage{dcolumn}
\usepackage{bm}
\usepackage{braket}
\usepackage{blkarray}
\usepackage{color}
\usepackage{appendix}
\usepackage{mathrsfs}

\begin{document}

\preprint{APS/123-QED}

\title{Generation of Robust Entanglement in Plasmonically Coupled Quantum Dots Driven by Quantum Squeezed Light}

\author{Sina Soleimanikahnoj}
\affiliation{The James Franck Institute, The University of Chicago, Chicago, Illinois 60637, USA}

\author{Stephen K. Gray}
\affiliation{Center for Nanoscale Materials, Argonne National Laboratory, Argonne, Illinois 60439, USA}

\author{Norbert F. Scherer}
\affiliation{Department of Chemistry and The James Franck Institute, The University of Chicago, Chicago, Illinois 60637, USA}

\date{\today}

\begin{abstract}
Our cavity quantum electrodynamics calculations demonstrate generation of steady-state entanglement between a plasmonically coupled pair of quantum dots by using single-mode squeezed light source. We show that strong coupling of plasmons to the incoming light source and the pairwise nature of squeezed photon generation enable the formation of entanglement between the initially unexcited quantum dots. The entanglement of quantum dots, measured as concurrence, can be improved replacing a pulsed source of light to continuous pumping of squeezed photons. Unlike previously introduced schemes the concurrence is robust against variations in the system parameters. Specifically, the generation of entanglement does not rely on fine tuning of plasmon quantum dot coupling. This work provides a new perspective for robust entangled state preparation in open quantum systems.
\end{abstract}

\maketitle


\section{Introduction}
Entanglement plays a central role in the fields of quantum computing and quantum information science in general~\cite{nielsen2002quantum,bouwmeester2000physics}. Generation of entanglement between quantum dots (QDs) has been the focus of much interest~\cite{gao2012observation,de2012quantum,bayer2001coupling,oliver2002electron}. Studies have shown that controlled interactions between two-level systems via a dissipative environment can create entanglement~\cite{poyatos1996quantum,lin2013dissipative,krauter2011entanglement}. Plasmonic nanostructures (PNs) placed in close proximity to QDs can provide the dissipative environment. In such hybrid plasmon-QD systems, the dissipative plasmons of the nanostructure (e.g., the plasmonic nanoparticle) can mediate the interaction of the QD excitations leading to creation of entanglement between the QDs~\cite{otten2019optical,otten2016origins,otten2015entanglement,lee2013robust,gonzalez2013non,he2012strong,martin2011dissipation,gonzalez2011entanglement}. In addition, PNs are capable of enhancing light-matter interactions by concentrating EM fields at the nanoscale. This enable the using a source of light to excite the plasmonic-QD structures and place QDs in an entangled state~\cite{otten2019optical,otten2016origins,otten2015entanglement}. However, such schemes of entanglement generation rely on fine tuning of plasmon-QDs coupling which requires precise positioning of the QDs with respect to the PN at the nanoscale~\cite{otten2015entanglement}.

 In the realm of cavity quantum electrodynamics, interaction of QD(s) with quantized modes of a cavity can lead to generation of entanglement between the QD and the cavity~\cite{ashhab2010qubit,zhou2020quantum}. However,in this situation entangled states are created in the ultra-strong coupling regime where QD-cavity coupling is comparable to QD and cavity resonance frequencies~\cite{ashhab2010qubit}. The ultra-strong coupling regime is extremely challenging to realize and experimental implementations are limited to a few specially designed architectures~\cite{bienfait2016reaching,petersson2012circuit,kubo2010strong,schuster2010high,pelton2015modified}. 
 Recent theoretical studies suggest squeezing the cavity mode by a parametric drive leads to an exponential increase of \textit{effective} QD-cavity coupling which leads to a transition from  weak to ultra-strong coupling regimes and consequent creation of entanglement~\cite{qin2018exponentially,leroux2018enhancing}. However, in the proposed schemes entanglement generation requires initializing the QD-cavity system to a specific quantum state or using a parametric drive with an extremely high quality factor.

In this paper, we present a method to generate entanglement between two QDs using a squeezed source of light, i.e. quantum light~\cite{loudon1987squeezed}. 
We use a PN to facilitate the absorption of incoming squeezed photons and mediate the interaction of the QDs. The scheme presented here does not require fine-tuning of the plasmon-QD coupling parameters. Steady-state entanglement can be generated from the ground state and there is no need for initialization of the system to a specific quantum state. Entanglement between the QDs is created in the weak coupling regime for conservative values of cavity quality factors ($Q \leq 200$).  Entanglement generation in dissipative quantum computing schemes often rely on intricate engineering of the dissipative environment of open quantum systems. The results presented here open new avenues for more robust entanglement generation in open quantum systems~\cite{verstraete2009quantum,kraus2008preparation}.

\section{Theoretical Methods}
\begin{figure}
\includegraphics[width= .6\linewidth]{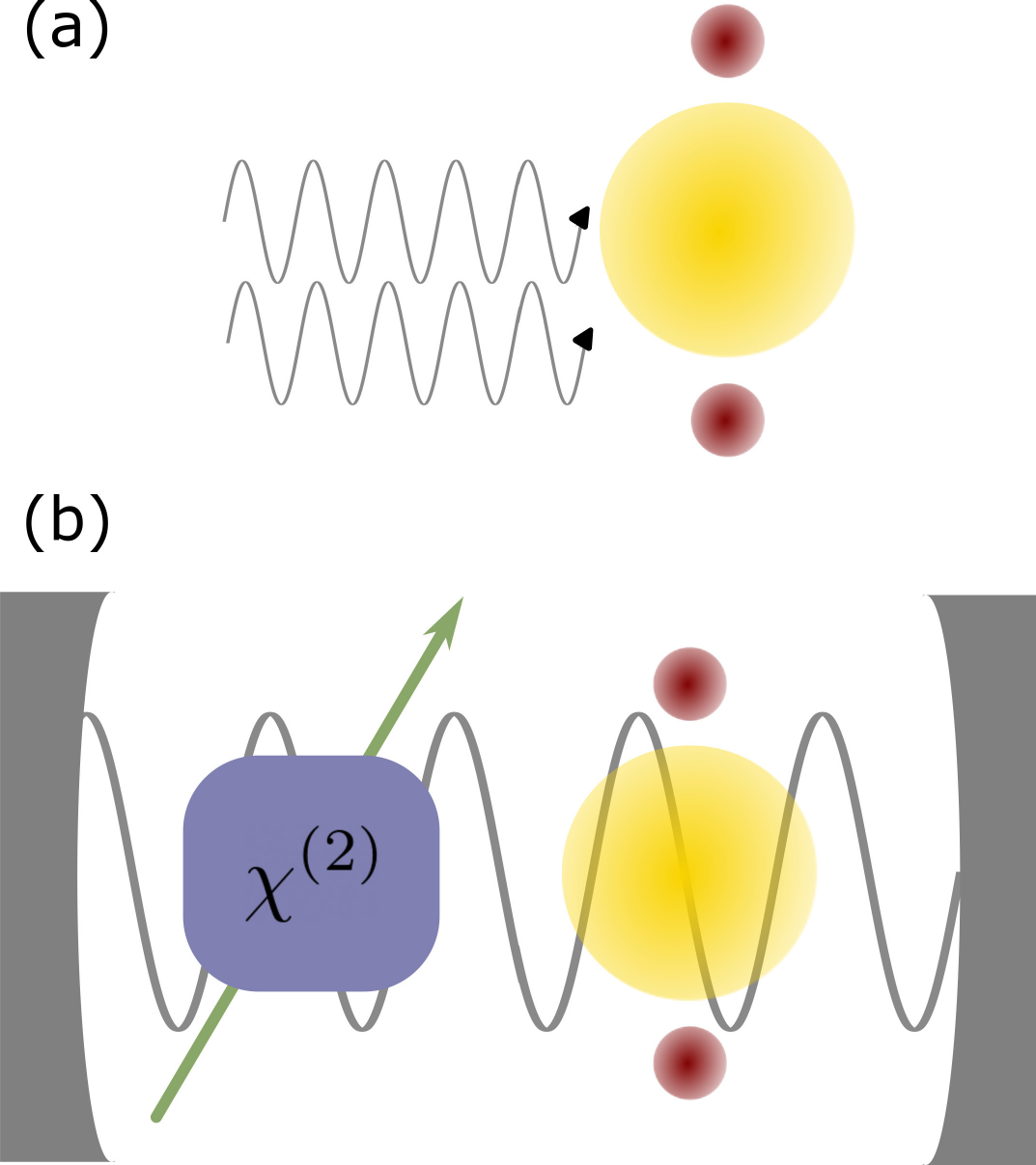}
\caption{Schematic of a system composed of a PN (shown in yellow) and two QDs (shown in red) in close proximity. In (a) the system is excited by a pulsed source of single-mode squeezed light, (e.g., colloquially understood as a  pair of entangled photons).  In (b), the system is placed inside a degenerate optical parametric oscillator that supplies a continuous source of squeezed light.}
\label{Fig:fig1}
\end{figure}

In our cavity quantum electrodynamics model 
we consider a system composed of a PN and a pair of quantum dots. The system is excited by a single-mode squeezed source of quantum light. The squeezed light is considered in two scenarios: as a single pulse (Fig. \ref{Fig:fig1}(a)) and in the continuous pumping limit where the system is embedded inside a cavity (Fig. \ref{Fig:fig1}(b)). Squeezed light can be produced by means of nonlinear optics~\cite{loudon1987squeezed}.  A bright pump field (green arrow) is focused into a nonlinear  $\chi^{(2)}$ crystal with strong second order susceptibility. Pump photons are down-converted into photons of half their frequency through careful phase matching of the waves involved. The PN (shown in yellow) strongly couples to the incoming squeezed photons and meditates the interaction between the two QDs (shown in red). 

In the rotating frame of squeezed photons the  Hamiltonian of the system in panel (b) is given by 
\begin{equation}
\begin{split}
&  H = \Delta_a \hat{a}^\dagger a + \Delta_b b^\dagger b + \epsilon(\hat{a}^\dagger \hat a^\dagger + \hat a\hat a) + g_{ab}(\hat a^\dagger \hat b+\hat b^\dagger \hat a)   \\ &  +\sum_{i=1}^2 \Delta_{c}\hat{c_i}^\dagger \hat c_i +
g_{bc}^i(\hat b^\dagger \hat c_i + \hat c_i^\dagger \hat b) + g_{ac}^i(\hat a^\dagger \hat c_i + \hat c_i^\dagger \hat a).
 \end{split}
 \label{Eq:7}
 \end{equation}
 Here, $\hat a(\hat a^\dagger)$, $\hat b(\hat b^\dagger)$ and $\hat c_i(\hat c_i^\dagger)$ are the annihilation (creation) operators for the squeezed photons, the plasmonic excitation of the nanostructure and the excitons in the quantum dot $i$, respectively. The $\epsilon(\hat{a}^\dagger \hat a^\dagger + \hat a\hat a)$ term in Eq.~\ref{Eq:7} captures the continuous pumping of the squeezed light \cite{leroux2018enhancing} and is omitted in single pulse simulations. The parameter $\epsilon$  is the parametric drive amplitude~\cite{leroux2018enhancing} which is proportional to the amplitude of the pump field in the parametric down conversion process~\cite{lvovsky2015squeezed}. We refer to this term as the driving amplitude. Parameters $g_{ab}$, $g_{bc}^i$ and $g_{ac}^i$ are coupling of photons and plasmons, plasmons and QD, and photons and QD excitations respectively. $\Delta_j = \omega_j - \omega_0$ where $ j=\hat a,\hat b,\hat c$  are detunings of frequency of cavity photons ($\omega_a$), plasmons ($\omega_b$) and QDs excitons ($\omega_{ci}$) with respect to driving squeezed field frequency ($\omega_0 = 2.04$ eV). For simplicity, we set all the detuning parameters to zero. While in certain cases this may render the solutions to parametric-type Hamiltonians unstable we have not faced any instability issues in our numerical results. This instability concern calls for a detailed analysis of Eq.~\ref{Eq:7}~\cite{zhang2021parity,leroux2018enhancing,gardiner2004quantum}. In the meantime, we attribute the stability of our solutions to the strong dampings of plasmons and photons since increased  damping has been shown to enhance the upper range of $\epsilon$ for a stable equation of motion~\cite{zhang2021parity}. Furthermore, we noticed the results remain unchanged under moderate cavity detuning ($|\Delta_a| \approx 2\epsilon$). we will show that significant cavity detuning $|\Delta_a| \gg 2\epsilon$ enables steady-state entanglement between the QDs. 
 
 The dipole moments of quantum emitters such as QDs are small ($1-10$ D)~\cite{hugall2018plasmonic}. As a result, direct coupling of photons and QDs is limited to the 0.1-2 meV range.
 We use photon-quantum dot coupling $g_{ac}^i = 2$ meV previously reported in the literature~\cite{shah2013ultrafast,otten2015entanglement}.
 PNs enable enhancement of light well within the sub-diffraction limit. The small mode volume of plasmonic systems enables huge coupling strengths (in the 1-100 meV range) to quantum emitters~\cite{shah2013ultrafast,pelton2015modified}. Unless stated otherwise, we use $g_{bc}^i = 50 $ meV through out the paper. Unlike quantum emitters, PNs couple strongly to the cavity photons~\cite{ameling2013microcavity}. According to Ref.~\cite{downing2017radiative}, at the diffraction limit,  plasmonic excitations of a metallic nanoparticles ($60$ nm radius) couple to photons with the strength of $\approx 111$ meV. We set the photon-plasmon coupling to $g_{ab} = 100$ meV. 
 \begin{table}[]
\begin{center}    
 \begin{tabular}{lll}
\multicolumn{2}{c}{} \\
\cline{1-3}
Parameter    & \qquad \qquad Value [meV]  & \qquad \qquad  Range [meV]\\
\hline
$g_{ab}$     & \qquad \qquad 100  & \qquad \qquad       0-200 \\
$g_{bc}$     & \qquad \qquad 50   & \qquad \qquad       20-80 \\
$g_{ac}$     & \qquad \qquad 2           \\
$\gamma_{a}$   & \qquad \qquad 10,40       \\
$\gamma_{b}$   & \qquad \qquad 150,50      \\
$\gamma_{c}$   & \qquad \qquad 1.7         \\
$\epsilon$      & \qquad \qquad 10    & \qquad \qquad    0-25    \\            
$\Delta_a$      & \qquad \qquad 0, 20, 50 \\
$\Delta_{b,c}$      & \qquad \qquad 0 \\

\hline
\end{tabular}
\end{center}
\label{table11}
\caption{Typical values and ranges of the parameters used in this paper. Range limits are separated by a dash line. Discrete values are separated by commas.}
\end{table}

The time evolution of the density matrix of the system $\hat \rho(t)$, is governed by the Lindblad master equation~\cite{otten2015entanglement}, 
\begin{equation}
\begin{gathered}
\frac{d \hat \rho(t)}{dt} = -\frac{i}{\hbar}\left[\hat H,\hat \rho\right] + \hat L\left(\rho\right).
\label{Eq:3}
 \end{gathered}
 \end{equation}

\noindent 
$\hat L(\hat \rho)  = \hat L_{\hat a}(\hat \rho)  + \hat L_{\hat b}(\hat \rho) + \hat L_{\hat c_1^+ \hat c_1}(\hat \rho) + L_{\hat c_2^+ \hat c_2}(\hat \rho)$ is the Lindblad superoperator. 
The first and the second terms incorporate the dissipation of the photonic and plasmonic excitations respectively.  The last two terms capture the dephasing of the excitons in quantum dots. 
Each term can be written as
\begin{equation}
\begin{gathered}
\hat L_{\hat x}(\hat \rho) = \frac{\gamma_{x}}{2}\left(2\hat x\hat \rho \hat x^\dagger - \hat x^\dagger \hat x \hat \rho - \hat \rho \hat x^\dagger \hat x \right).
 \end{gathered}
 \end{equation}
 For photons in the cavity damping is set to $\gamma_{a} = 10$ meV ($Q \approx 200$), which is well within the experimentally feasible range of squeezed light sources~\cite{marty2021photonic,hu2016design}. The damping of PNs is $\gamma_{b} =  150$ meV and the dephasing energy parameter of QDs is chosen to be $\gamma_{c} = 1.7$ meV~\cite{otten2015entanglement,otten2016origins,shah2013ultrafast}.  Since, $g_{ab}, g_{bc}^i <\gamma_{b}$ and $g_{ac}^i <\gamma_{a}$ our study is in the weak coupling regime.
Table~I summarizes the range and typical values of parameters used throughout the paper.  
 
One can use Wootters' concurrence \cite{Wootters1998} to benchmark the entanglement between the QDs. This concurrence, which is 0 for separable and 1 for maximally entangled states, is inferred from the eigenvalues of a ``spin-flipped'' (and complex-conjugated) bipartite density matrix.  Concurrence is calculated according to the procedures explained in Appendix A. The system is studied using the basis states spanned by the excitation number of photons ($n_{a})$, plasmons ($n_{b}$) and quantum dots 1 and 2 ($n_{c1},n_{c2}$): $\ket{n_{a},n_{b},n_{c1},n_{c2}}$. The state of the quantum dots is given in terms of the Bell states:
\begin{subequations}
\begin{equation}
\ket{B_1} = \sqrt{\frac{1}{2}}\left(\ket{0,0} + \ket{1,1}\right) 
\end{equation} 
\begin{equation}
\ket{B_2} = \sqrt{\frac{1}{2}}\left(\ket{0,0} - \ket{1,1}\right) 
\end{equation}
\begin{equation}
\ket{B_3} = \sqrt{\frac{1}{2}}\left(\ket{0,1} + \ket{1,0}\right) 
\end{equation}
\begin{equation}
\ket{B_4} = \sqrt{\frac{1}{2}}\left(\ket{0,1} - \ket{1,0}\right).
\end{equation}
\label{bell}
\end{subequations}
\noindent States $\ket{B_1}$, $\ket{B_2}$, $\ket{B_3}$, $\ket{B_4}$ are often
labelled $\ket{\Phi^+}, \ket{\Phi^-}$, $\ket{\Psi^+}$, $\ket{\Psi^-}$, respectively.
Each Bell state is maximally entangled and together the four Bell states form a complete basis for the Hilbert space of the two quantum dots. This allows writing the state of the system in the basis: $\ket{n_{a},n_{b},B_{1-4}}$. An intuitive understanding of how entanglement is obtained by the studying the population of the Bell states. Similar to previous studies \cite{otten2015entanglement,gonzalez2011entanglement}, we show that the difference in their population correlates with concurrence. The population of the Bell states $B_i$, $i = {1-4}$  are found by calculating 
the trace of $\hat{\rho} ~ \hat{I}_{ph} \otimes \hat{I}_{pl} \otimes \ket {B_i}\bra{B_i}$ where $\hat I_{ph}$ and $\hat I_{pl}$ are identity matrix operators for photons and plasmons, respectively. The density matrix is found by solving Eq.~\ref{Eq:3} for $t>0$ using the Quantum Toolbox in Python QuTiP~\cite{johansson2012qutip,JOHANSSON20131234}.

\section{Results}
\subsection{Single Pulse Case}

\begin{figure}[]
	\centering
	\includegraphics[width= .6\linewidth]{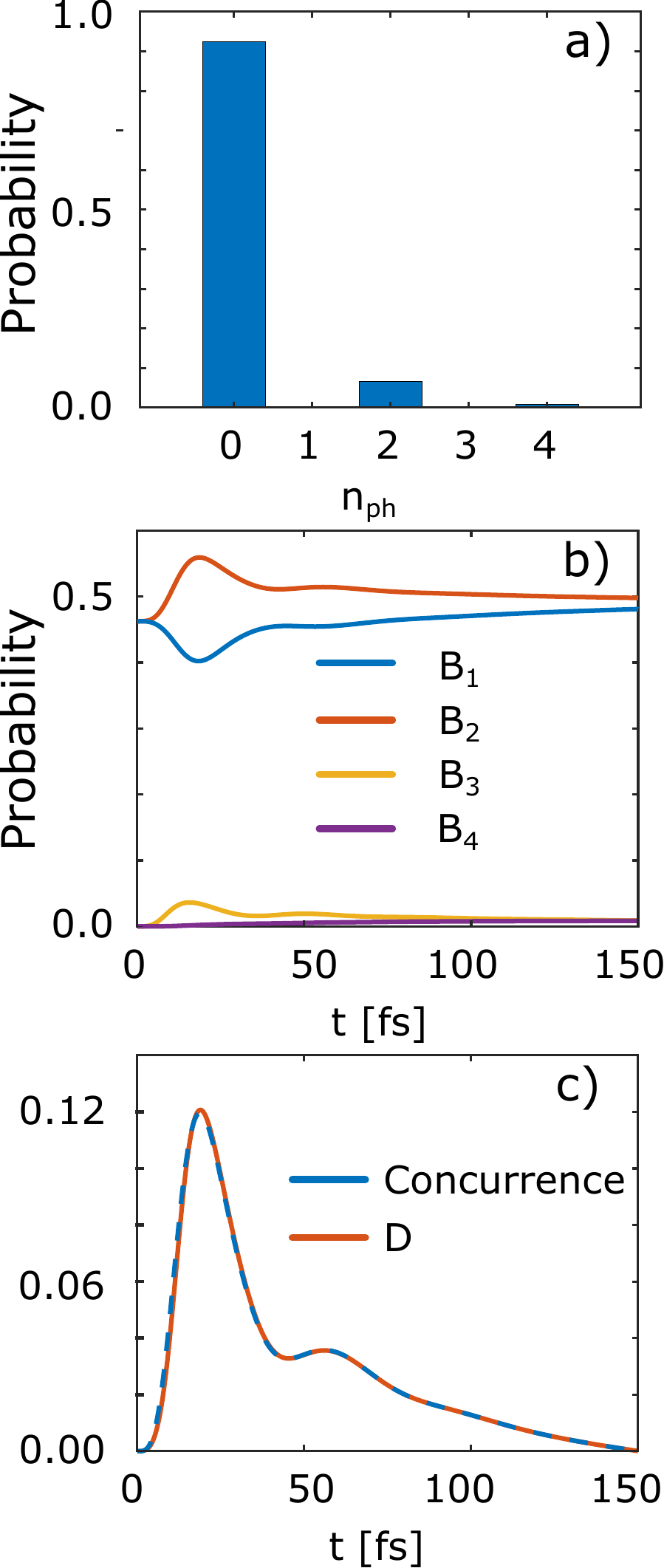}
    \caption{(a) Probability of photonic number states for a squeezed pulse of light given by Eq.~\ref{Eq:1}. $Z = 0.2$. The probability of states with $n_{a} \geq 4$ is negligible and the photonic ket can be approximated by Eq.~\ref{Eq:1}. (b) and (c) are density matrix results for populations of the Bell states, concurrence and $D$ as a function of time for a squeezed pulse with the number state probabilities shown in panel (a).}
    \label{Fig:fig2}
\end{figure}

In the single pulse simulations the photon squeezing strength is parameterized by the complex number $Z = r^{i\theta}$ where $r$ and $\theta$ denote the strength and phase of squeezing, respectively. Squeezed photons are prepared by applying the squeezing operator $S(Z) = \exp\left({Za^\dagger a^\dagger - Z^*aa}\right)$ to the vacuum states of photons $\ket{0_{ph}}$~\cite{lvovsky2015squeezed}:
\begin{equation}
\begin{split}
\ket{\xi_0} &  = S(Z)\ket{0}_{ph} \\ &=   \frac{1}{\sqrt{\cosh(r)}}\sum_{n=0}^{\infty}\frac{\sqrt{2n!}}{n!}\left\{-\frac{1}{2}e^{i\theta} \tanh(r)\right\}^n\ket{2n}.
\label{Eq:1}
\end{split}
\end{equation}
The squeezed single-mode photon state (Eq.~\ref{Eq:1}) involves only even photon numbers. This fundamental property of this state follows from pairwise appearance of creation and annihilation operators in $S(Z)$.
Fig.~\ref{Fig:fig2}(a) shows the probability of those states for $Z = 0.2$, the only states with non-negligible probabilities are $n_{a} = 2n = 2$ and the vacuum state. 
Therefore this squeezed state can be approximated as 
\begin{equation}
\begin{split}
\ket{\xi_0}  \approx \frac{1}{\sqrt{\cosh(r)}}\left\{\ket{0}_{ph}  -\frac{\tanh(r)}{\sqrt{2}}\ket{2}_{ph}\right\}.
\label{Eq:2}
\end{split}
\end{equation}
We use Eq.~\ref{Eq:1} to initialize the photons in a squeezed state. Assuming the PN and the quantum dots are in the ground state at time zero, the initial state of the entire system is 
$\rho(0) = \ket{\xi_0,0,0,0}\bra{\xi_0,0,0,0}$. 
Figure~\ref{Fig:fig2}(b) shows the population of the Bell states as a function of time. Bell states $B_1$ and $B_2$, which are the superposition of kets with even number of total excitations ($\ket{0,0}$ and $\ket{1,1}$), have the more pronounced changes in their population compared to $B_3$ and $B_4$. The latter combinations of kets with an odd number of total excitations ($\ket{0,1}$ and $\ket{1,0}$). We define the parameter $D$ as
\begin{equation}
\begin{split}
D = \max \left\{ P_{B_2} - \left(P_{B_1} + P_{B_3} + P_{B_4}\right),0\right\},
\end{split}
\end{equation}
\noindent to benchmark the population difference of the Bell states. 
Specifically, $D$ is a measure that compares the probability of the most populated Bell state (in this case $B_2$) to the population of the other three Bell states.
Figure~\ref{Fig:fig2}(c) shows $D$ and concurrence as a function of time. Concurrence mirrors $D$ both peaking around $20$ fs and vanishing at $150$ fs. Therefore, entanglement characterized by concurrence corresponds to the difference in population of the Bell states given by $D$, which is a
more intuitive quantity.
Still, we need to clarify what gives rise to this population difference of the Bell states in the first place. 
The mechanism  which causes this population difference can be understood by solving the Schr{\"o}dinger equation for the non-Hermitian Hamiltonian model of the system. In this model, dephasing and dissipation of excitations are incorporated as imaginary diagonal terms in the Hamiltonian which we refer to as $H_{int}$, and the time-dependent Schrödinger equation is solved using this modified Hamiltonian~\cite{cortes2020non}. While the non-Hermitian model is an approximation, it can effectively emulate the qualitative (and sometimes quantitative) behavior of the system with significantly lower computational cost compared to the Lindblad master equation. Here, this model is primarily used for building intuition and understanding  the entanglement generation mechanism. For this system, $H_{int}$ is
\begin{equation}
\begin{split}
H_{int}& =  \sum_{i=1}^2-i\gamma_{a}\hat a^\dagger \hat a   - i\gamma_{b}\hat b^\dagger \hat b  -i\gamma_{c} \hat c_i^\dagger  \hat c_i  \\ &  +g_{ab}(\hat a^\dagger \hat b+ \hat a \hat b^\dagger) + g_{bc}^i(\hat b^\dagger \hat c_i + \hat c_i^\dagger \hat a)~.
\label{Eq:6}
\end{split}
\end{equation}
The coupling between photons and quantum dots ($g_{ac}^i$) is weak and has negligible effect on the population of states and concurrence. Therefore, it is ignored in the non-Hermitian Hamiltonian.  The ket describing the system can be expanded, as shown in  Appendix B,
\begin{equation}
\begin{split}
\ket{\psi(t)} & = 
\alpha_1(t)\ket{0,0,B_1} + 
\alpha_1(t)\ket{0,0,B_2} + 
\alpha_3(t)\ket{0,1,B_3}  \\
 & +  \alpha_4(t)\ket{1,0,B_3} +
\alpha_5(t)\ket{1,1,0,0} +
\alpha_6(t)\ket{2,0,0,0}.
\label{Eq:4}
\end{split}
\end{equation}

 We use Eq.~\ref{Eq:2}  to initialize the system  as $\ket{\psi(0)} = \ket{\xi_0,0,0,0}$. Only kets with total excitation number of $2$ ($n_{tot} = n_{a} +n_{b} + n_{c1} + n_{c2} = 2$) are considered in Eq.~\ref{Eq:4} for two reasons: First, according Eq.~\ref{Eq:2}, squeezed photons are generated in pairs and states with excitation numbers other than $2$ do not couple to the squeezed photons ($\braket{\psi_{n_{tot}\neq2}|H_{int}|2,0,0,0} = 0$). Second, states with different excitation numbers do not couple to each other ($\braket{\psi_{n_{tot}}|H_{int}|\psi_{m_{tot}}}=0, m_{tot}\neq n_{tot}$). Therefore, the probability amplitude of states with $n_{tot} \neq 2$ remains zero for $t>0$ and can be ignored in Eq.~\ref{Eq:4}. 
\begin{widetext}
\begin{gather}
\label{Eq:5}
i\hbar \begin{bmatrix}
\dot \alpha_1 \\
\dot \alpha_2 \\
\dot \alpha_3 \\
\dot \alpha_4 \\
\dot \alpha_5 \\
\dot \alpha_6 \\
\end{bmatrix}
=
\begin{blockarray}{ccccccc}
\begin{block}{(cccccc)c}
  -i2\gamma_{c} & 0 & +g_{bc} & 0 & 0 & 0 &   \\
  0 & -i2\gamma_{c} & -g_{bc} & 0 & 0 & 0 &   \\
  +g_{bc} & -g_{bc} & -i(\gamma_{b}+\gamma_{c})  & g_{ab} & 0 & 0 &  \\
  0 & 0 & g_{ab} & -i(\gamma_{a}+\gamma_{c})  & g_{bc} & 0 &  \\
  0 & 0 & 0 & g_{bc} & -i(\gamma_{a} + \gamma_{b}) & \sqrt{2}g_{ab} &  \\
  0 & 0 & 0 & 0  & \sqrt{2}g_{ab} &-i2\gamma_{a} &  \\
\end{block}
\end{blockarray}
  \begin{bmatrix}
 \alpha_1 \\
 \alpha_2 \\
 \alpha_3 \\
 \alpha_4 \\
 \alpha_5 \\
 \alpha_6
\end{bmatrix}
\end{gather}
\end{widetext}
Also, we assume the couplings of the the QDs with the PN are equal ($g_{bc}^1 = g_{bc}^2$). As a result, the photonic and plasmonic antisymmetric states  ($\ket{1,0,B_4}$ and $\ket{0,1,B_4}$) do not couple to the rest of the states in Eq.~\ref{Eq:4} and are ignored. This is in agreement with the density matrix results shown in Fig~\ref{Fig:fig2}(b) where the antisymmetric state, $B_4$ has negligible population.
In order to find the probability amplitude of the states in Eq.~\ref{Eq:4} we write a Schr{\"o}dinger equation for the non-Hermitian Hamiltonian in Eq.~\ref{Eq:6}.
This results in a linear set of first-order differential equations [See Eq.~\ref{Eq:5}] which we solve numerically.
\begin{figure}[]
	\centering
	\includegraphics[width= .6\linewidth]{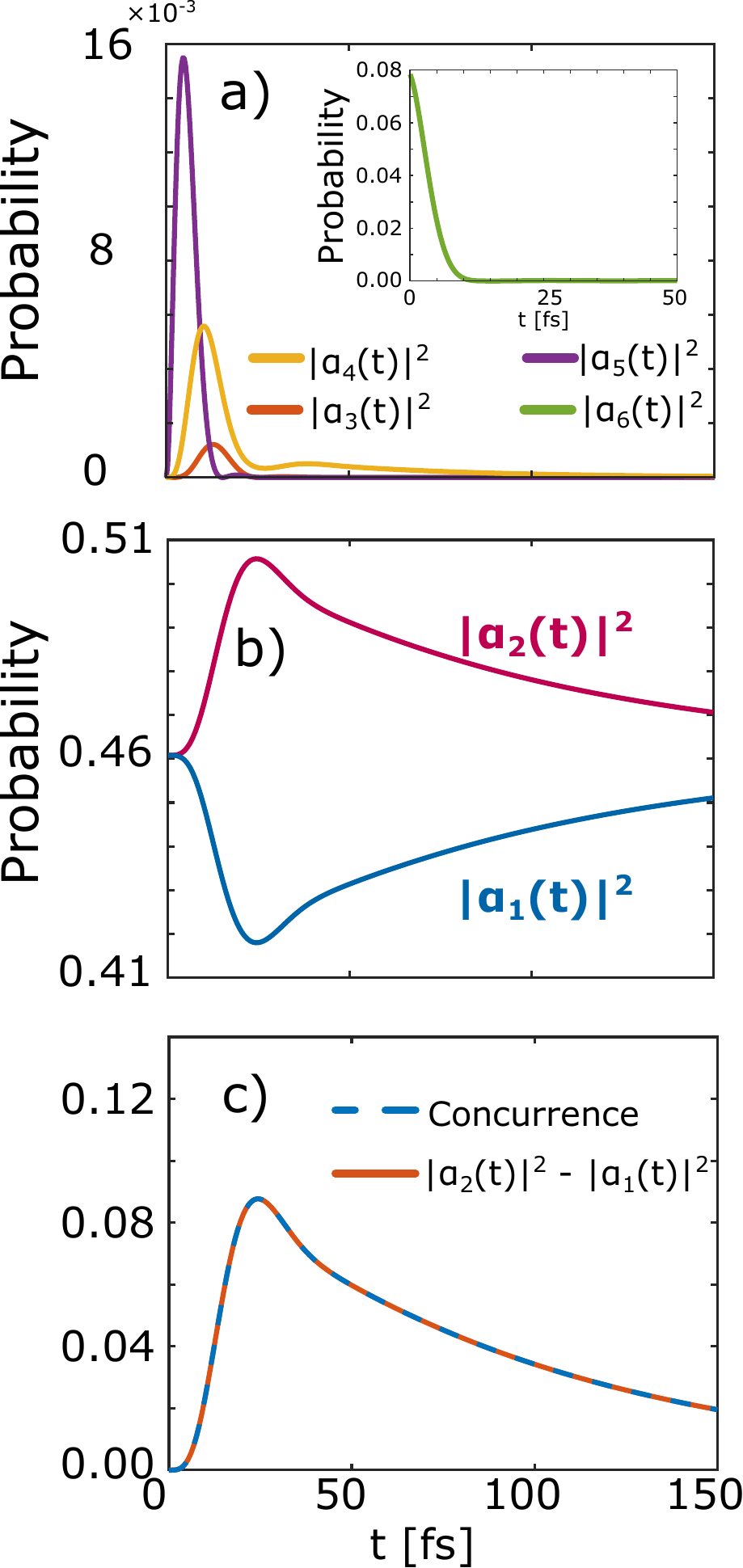}
    \caption{States probabilities and concurrence  when a plasmon-two-QD system initially is exposed to a squeezed photon pulse (Eq~\ref{Eq:1}). The Schr{\"o}dinger equation is solved for the non-Hermitian Hamiltonian in Eq.~ \ref{Eq:2} for the basis given in Eq.~\ref{Eq:4}. (a),(b) show the various state probabilities, and (c) displays concurrence and population difference of the Bell states as a function of time.  }
    \label{Fig:fig3}
\end{figure}

Figure~\ref{Fig:fig3}(a) shows the probability of each of the states as a function of time. Panel (b) shows $|(\alpha_1 (t)|^2$ (blue) and $|\alpha_2 (t)|^2$ (magenta) which are the  populations of the states $\ket{0,0,B_1}$ and $\ket{0,0,B_2}$ respectively.
\begin{figure}
	\centering
	\includegraphics[width= .6\linewidth]{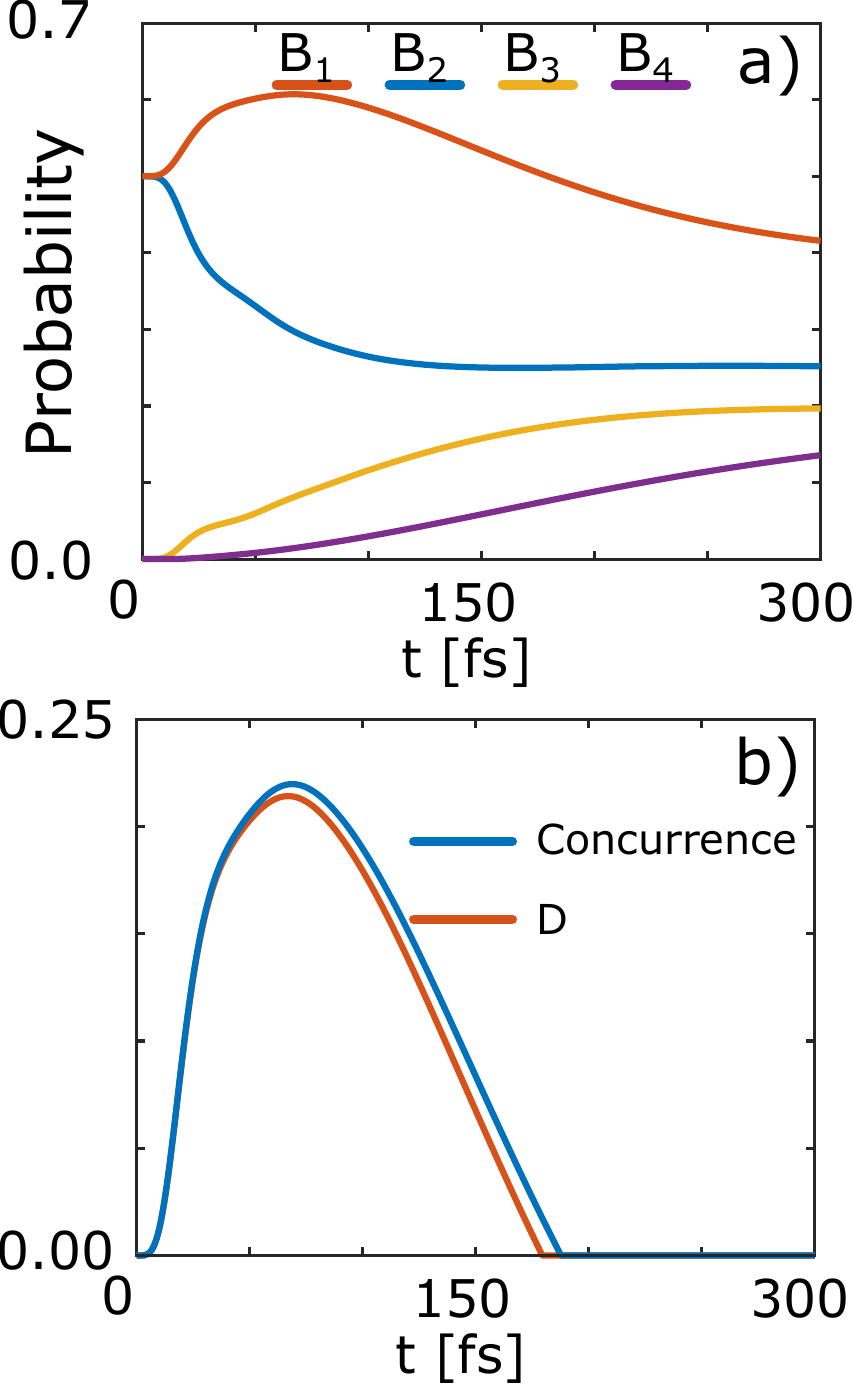}
    \caption{Bell states probabilities and concurrence when a plasmon-two-QD system initially in the ground state is placed inside a degenerate parametric oscillator. (a) Probability of Bell states (a), concurrence and population difference of the Bell states (b) as a function of time. }
    \label{Fig:fig4}
\end{figure}
As expected, $|\alpha_6 (t)|^2$ peaks at time zero and the initial excitation is in the two-photon state ($\ket{2,0,0,0}$).  Due to the tridiagonal structure of the Hamiltonian matrix in Eq.\ref{Eq:5}, the excitation cascades through states $\ket{1,1,0,0}$, $\ket{1,0,B_3}$ and $\ket{0,1,B_3}$ progressively. In Fig.~\ref{Fig:fig3}(a),(b) this is highlighted by successive appearance of the peaks of the corresponding probabilities of these states. As shown in Fig.~\ref{Fig:fig3}(b), the squeezed photons populate $\ket{0,0,B_2}$ and depopulate the state $\ket{0,0,B_1}$. This is in agreement with the findings of the density matrix calculations shown in Fig.~\ref{Fig:fig2}(b). Also, similar to the density matrix results this population/depopulation of the Bell states is what gives rise to entanglement.  Fig.~\ref{Fig:fig3}(c) shows that the population difference of the Bell states $\ket{0,0,B_2}$  and $\ket{0,0,B_1}$  is in close agreement with concurrence.

The mechanism behind the population and depopulation of the Bell states can be understood by considering Eq.~\ref{Eq:5}. The coupling of the plasmonic symmetric state ($\ket{0,1,B_3}$)  to $\ket{0,0,B_1}$ is $+g_{bc}$ which is out phase with respect to its coupling to $\ket{0,0,B_2}$ ($\braket{0,1,B_3|H_{int}|0,0,B_2} = -g_{bc}$). As a result, $\ket{0,0,B_1}$ gets populated and the state $\ket{0,0,B_2}$ depopulates.

\begin{figure}[]
	\centering
	\includegraphics[width= .6\linewidth]{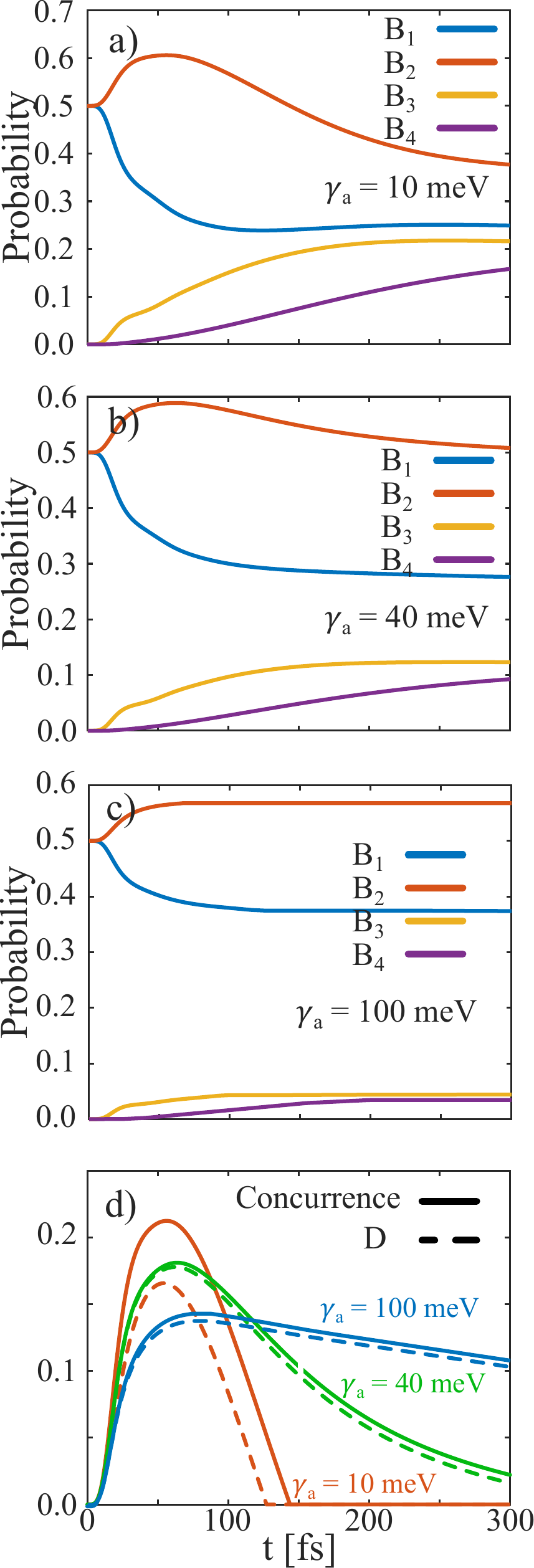}
    \caption{Continuous pumping density matrix results for effect of cavity damping on entanglement. Population of the Bell states for photonic damping of 10 meV (a), 40 meV(b) and 100 meV. (c) Concurrence and population difference of the Bell states ($D$) for the three different photonic dampings.}
    \label{Fig:fig5}
\end{figure}
\subsection{Continuous Pumping Results}
So far, we explored a configuration where squeezed photons were considered as an initial excitation pulse. Alternatively, one can imagine placing the plasmon-QDs structure inside a degenerate parametric oscillator where a continuous ray of squeezed photons is supplied.as depicted in Fig.~\ref{Fig:fig1}(b). The Hamiltonian describing this continuous pumping case is given by Eq.~\ref{Eq:7}. We assume that the constituents of the system are initially in their ground states. Similar to the single-pulse simulations, the populations of the states and concurrence are extracted from the the density matrix  which is found by solving Eq.~\ref{Eq:3}. 

Figure~\ref{Fig:fig4}(a) shows the populations of the Bell states as a function of time. Unlike the single pulse simulation, here the population of the antisymmetric Bell state $B_4$ is nonzero. This is due to incoherent coupling of the $B_4$ state  to other states via Lindbladian terms of the master equation~\cite{gardiner2004quantum}. These terms were ignored in the non-Hermitian Hamiltonian formalism, as photonic and plasmonic antisymmetric states do not coherently couple to the rest of the states in Eq.~\ref{Eq:4}.

In analogy to the single pulse case, couplings of the $B_1$ and $B_2$ states to the plasmonic symmetric state are out of phase, which causes depopulation of the former and population of the latter state. As shown in Fig.~\ref{Fig:fig4}(b) the population difference of the Bell states gives rise to entanglement measured as concurrence. In comparison with single pulse simulation results (Fig.~\ref{Fig:fig2}), the peak concurrence shows a two fold increase.

The population changes of states $B_1$ and $B_2$ are larger in comparison to states $B_3$ and $B_4$.
This can be understood by revisiting Eq.~\ref{Eq:5}. A diagonal element of the matrix determines the damping rate of its corresponding state. While the damping of states $\ket{0,0,B_1}$ and $\ket{0,0,B_2}$  are $2\gamma_{c}$, plasmonic and photonic symmetric states ($\ket{0,1,B_3}$ and $\ket{1,0,B_3}$) are damped at much stronger rates dictated by the damping of the PN or the photonic cavity respectively ($\braket{0,1,B_3|H_{int}|0,1,B_3} \approx \gamma_{b}$ and $\braket{1,0,B_3|H_{int}|1,0,B_3} \approx \gamma_{a}$). A similar argument applies to state $B_4$. According to Eq.~\ref{Eq:6}, the plasmonic and photonic antisymmetric states are damped at $\braket{0,1,B_4|H_{int}|0,1,B_4} \approx \gamma_{b}$ and $\braket{1,0,B_4|H_{int}|1,0,B_4} \approx \gamma_{a}$, respectively. This is considerably stronger than the dissipation of $B_{1,2}$    ($\braket{0,0,B_{1,2}|H{int}|0,0,B_{1,2}} =2\gamma_{c}$).

The effect of damping can be further understood by exploring the results for three different values of cavity dampings. Figs.~\ref{Fig:fig5}(a) , (b) and (c) show the populations of the Bell states as a function of time for cavity damping of $\gamma_{a} = 10$ meV, $\gamma_{a} = 40$ meV and  $\gamma_{a} = 100$ meV,  respectively. Concurrence and $D$ are shown in Fig~\ref{Fig:fig5}(d).
Since $\braket{1,0,B_3|H_{int}|1,0,B_3} \approx \gamma_{a}$ and $\braket{1,0,B_4|H_{int}|1,0,B_4} \approx \gamma_{a}$  increase of cavity  damping  leads to increased  damping of photonic symmetric and antisymmetric states. As a result, $B_3$ and $B_4$ are populated more slowly and their steady state populations have decreased. According to panel (d), for $\gamma_{a} = 100$ meV this decrease of populations makes $D$ nonzero at steady-state and facilities developing steady-state entanglement.
\begin{figure}[]
	\centering
	\includegraphics[width= .7\linewidth]{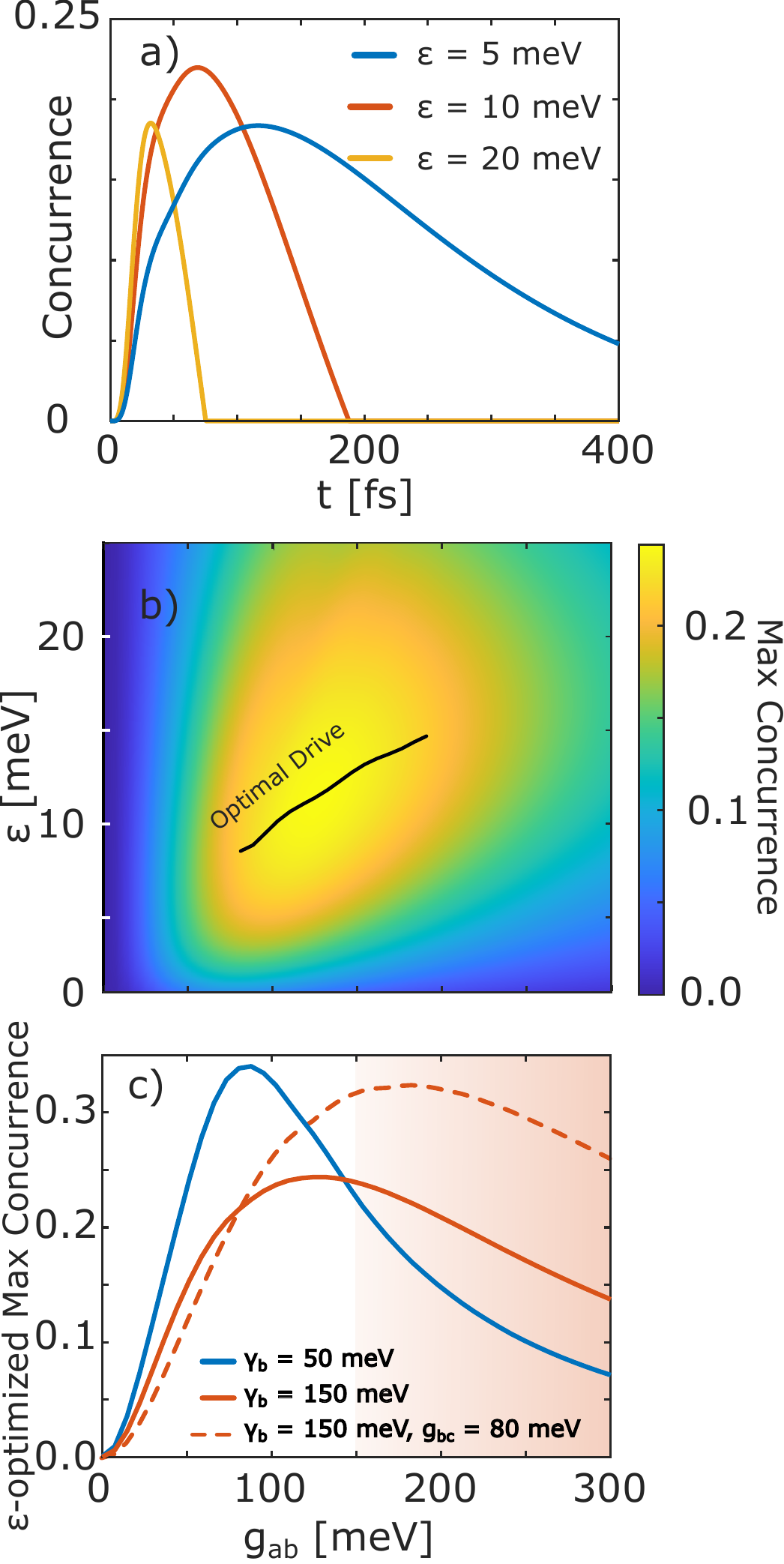}
    \caption{Probabilities of Bell state and concurrence for values of driving amplitude that maximize concurrence. (a) Concurrence for driving amplitude of 5 meV, 10 meV and 20 meV. (b) Maximum concurrence of QDs as determined by 930 simulations as a function of photon-plasmon coupling ($g_{ab}$)and driving amplitude ($\epsilon$) For panels (a) and (b) $\gamma_{b} = 150$ meV and $g_{bc} = 50$ meV. The black curve represents the optimum $\epsilon$ for a give $g_{ab}$. (c) $\epsilon$-optimized max concurrence as a function of photon-plasmon coupling for three different sets of simulation parameters. For both solid curves $g_{bc} = 50$ meV. The shaded region highlights the  $g_{bc} > \gamma_{b}$ regime for the two red curves.}
    \label{Fig:fig6}
\end{figure}

One other parameter that determines the populations of the Bell states and concurrence is the driving amplitude, $\epsilon$. It is shown in  Fig.~\ref{Fig:fig6}(a) that for a fixed set of coupling parameters there is an optimum value of $\epsilon$ which gives maximum concurrence.  At $\epsilon = 5$ meV, the population difference of $B_1$ and $B_2$ is small which results in a smaller peak concurrence compared to the concurrence at $\epsilon = 10$ meV. Also, increasing the driving amplitude from $10$ to $20$ meV decreases peak concurrence. This is due to an increase in the rate of population of $B_3$ and $B_4$ which leads to a smaller population difference of Bell states and peak concurrence. Figure~\ref{Fig:fig6}(b) shows maximum concurrence for 930 simulations as a function of driving amplitude and photon-plasmon coupling. For each value of coupling  there is an optimum value of  driving amplitude that gives maximum concurrence depending on the population difference of the Bell states. Furthermore, Fig.~\ref{Fig:fig6}(b) highlights the critical role of the PN; a significant concurrence is only achieved for values of $g_{ab} \geq 60$ meV. Strong coupling to photons is a characteristic of PNs which often comes at the price of strong damping. The black line in panel (b) shows the optimal $\epsilon$ for which maximum value of peak concurrence is achieved as a function of $g_{ab}$.

The $\epsilon$-optimized max concurrence is plotted as a function $g_{ab}$ in Fig.~\ref{Fig:fig6}(c) for three different sets of simulation parameters. In the strong coupling regime where  $g_{ab} > \gamma_{b}$, optimized concurrence decreases with the increase of photon-plasmon coupling. For simulations with $\gamma_{b} = 150$ meV  this region is highlighted in red. The reason for this decrease of optimized concurrence is not understood and requires further study of the strong coupling regime. 
Fig.~\ref{Fig:fig6}(c) highlights an advantage of using a more lossy PN. Increase of $\gamma_{b}$  widens the energy range of photon-plasmon coupling where a significant value of concurrence can be achieved. In addition, in the weak photon-plasmon coupling regime optimized concurrence can be further improved by increasing the plasmon-QD coupling. The strong interaction of QDs and plasmons is feasible and has been achieved in experiment for quantum emitters embedded inside plasmonic cavities~\cite{santhosh2016vacuum,bitton2019quantum,leng2018strong}. 
\begin{figure}[]
	\centering
	\includegraphics[width= .6\linewidth]{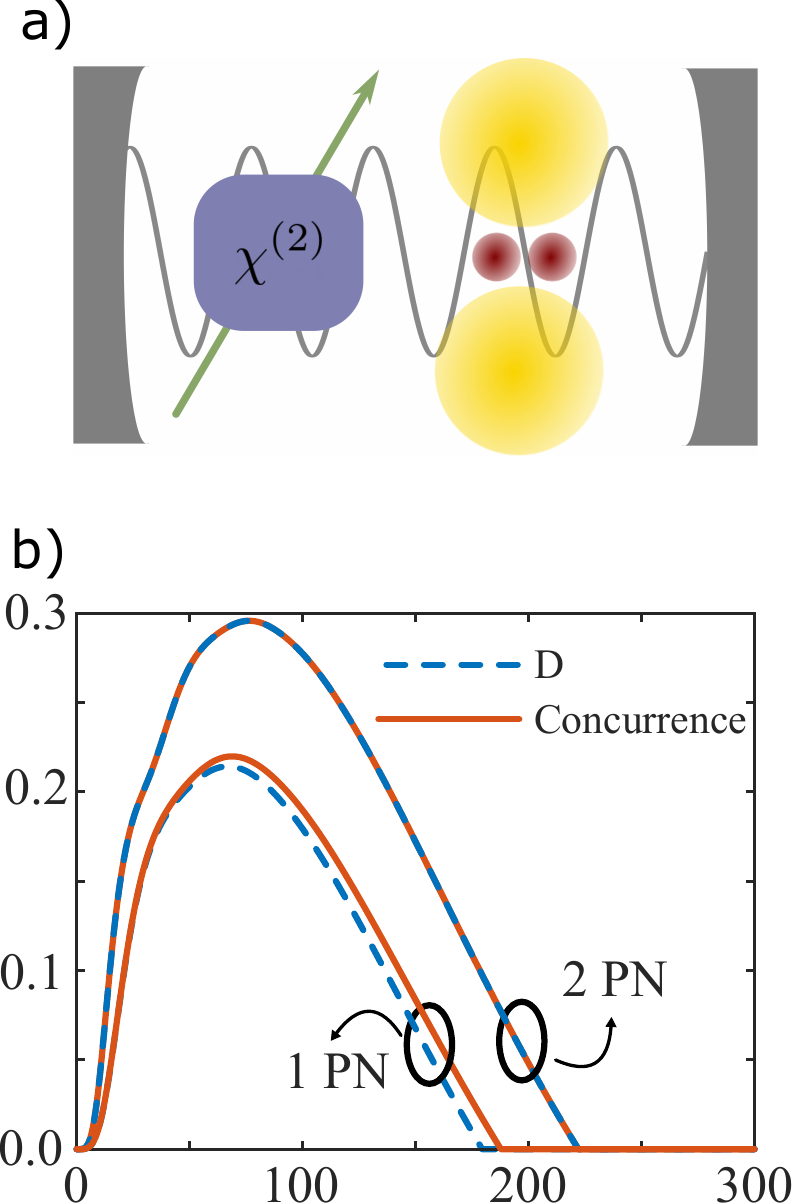}
    \caption{The effect of the number of PN on $D$ and concurrence. (a) Schematic of the system studied. A high intensity pump field (green arrow) is incident upon a nonlinear $\chi^{(2)}$ material. The down conversion process leads to creation of pairs of photons with half the frequency of the pump field.  Yellow (red) circles represent the PNs (QDs). (b) Concurrence and $D$  as function of time for 1 or 2 PNs.}
    \label{Fig:fig7}
\end{figure}
\begin{figure}
	\includegraphics[width= .6\linewidth]{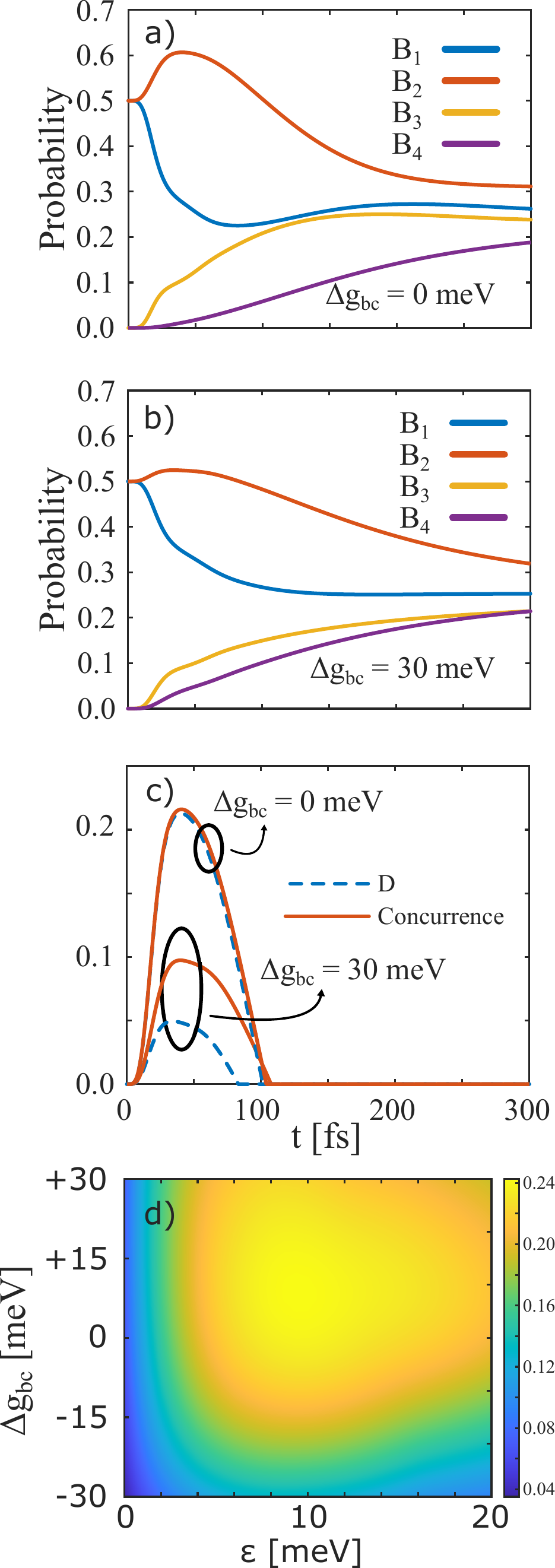}
    \caption{The effect of NP-QD coupling mismatch on entanglement. (a) Populations of Bell states when there is no coupling mismatch. (b) Populations of Bell states for $\Delta g_{bc} = 30 $ meV. (c) Concurrence and $D$ in presence and absence of NP-QD coupling mismatch. Maximum concurrence for 930 simulations as a function driving amplitude and $\Delta g_{bc}$.}
    \label{Fig:fig8}
\end{figure}

The maximum value of the concurrence and its range can be improved with the addition of another PN. This addition leads to further concentration of light at the nanoscale~\cite{shah2013ultrafast}. The Hamiltonian and the Lindblad equation for this configuration are discussed in Appendix C and the schematic is shown in Fig.~\ref{Fig:fig7}(a). Concurrence and $D$ are shown in Fig.~\ref{Fig:fig7}(b).  The second PN modifies Eq.~\ref{Eq:4} with an extra pair of photonic and plasmonic symmetric states which provides additional coupling to the excitation cascading from the two photon state to the Bell states $B_1$ and $B_2$, and leads to a larger population difference of $B_1$ and $B_2$ and larger concurrence.

The plasmon-QD couplings ($g_{bc}^i$) used in this manuscript are well within the feasible range. However, $g_{bc}^1$ and $g_{bc}^2$ depend on parameters such as the distance between each of the the QDs and the PN, which are difficult to control in an experiment. As a result, there will be an inevitable mismatch between the coupling of the quantum dots and the NP ($\Delta g_{bc} = g_{bc}^1 - g_{bc}^2 \neq 0$). The effect of this mismatch on concurrence is explored in Fig.~\ref{Fig:fig8}. Panels (a) and (b) show the population of the Bell states for $\Delta g_{bc} = 0 $ and $\Delta g_{bc} = 30 $ meV respectively. In presence of a mismatch, the populations of Bell states $B_2$ and $B_3$ decrease and those of $B_1$ and $B_4$ increase. This leads to the decrease of $D$ and concurrence shown in Fig.~\ref{Fig:fig8}(c).

The changes in these populations and concurrence can be explained as follows: When $\Delta g_{bc} $ is nonzero, the magnitude of coupling between the plasmonic symmetric states and Bell states $B_1$ and $B_2$ decreases ($\braket{0,0,B_{1-2}|H_{int}|0,1,B_3} = \pm g_{bc} \mp \frac{\Delta g_{bc}}{2}$) which results in the decrease and increase of population Bell states $B_2$ and $B_1$, respectively. Also, the coupling mismatch coherently couples the photonic and plasmonic antisymmetric states ($\ket{1,0,B_4}$ and $\ket{0,1,B_4}$) to the rest of two-excitation states in Eq.~\ref{Eq:4}, which explains the increase of the population of $B_4$ shown in panel (b). Furthermore, the mismatch weakens the coupling of the plasmonic and photonic symmetric states ($\ket{1,0,B_3}$ and $\ket{0,1,B_3}$) to the rest of two-excitation states. This decreases the population of $B_3$. 

It should be noted that a deliberately large of value  of $\Delta g_{bc} = 30$ meV is chosen to emphasize the effect of mismatch on concurrence. Despite the results in Fig.~\ref{Fig:fig8}(c), it can be argued that the entanglement generation scheme presented here is relatively robust against PN-QD coupling mismatch. Figure~\ref{Fig:fig8} plots maximum value of concurrence as a function of driving amplitude and PN-QD coupling mismatch. As shown, for moderate values of driving amplitude ($\epsilon \approx 10$ meV) significant concurrence can be achieved for $\Delta g_{bc} \geq -15$ meV. This is a stark improvement over previously introduced schemes~\cite{otten2015entanglement}. In earlier works in order to generate entanglement in QD-PN systems, a classical source of light was used.  When the incoming field is classical, fine tuning of $\Delta g_{bc}$ to an optimum value is required to create concurrence. In these schemes, a slight change of $\pm 5$ meV in the optimum value of $\Delta g_{bc}$ causes the maximum concurrence to go from $\approx 0.25$ to zero. 
By contrast, in the scheme presented here, the only parameter that requires fine tuning is $\epsilon$ which can be easily controlled by tuning the pump field intensity in the parametric down-conversion process.
\begin{figure}
	\includegraphics[width= .6\linewidth]{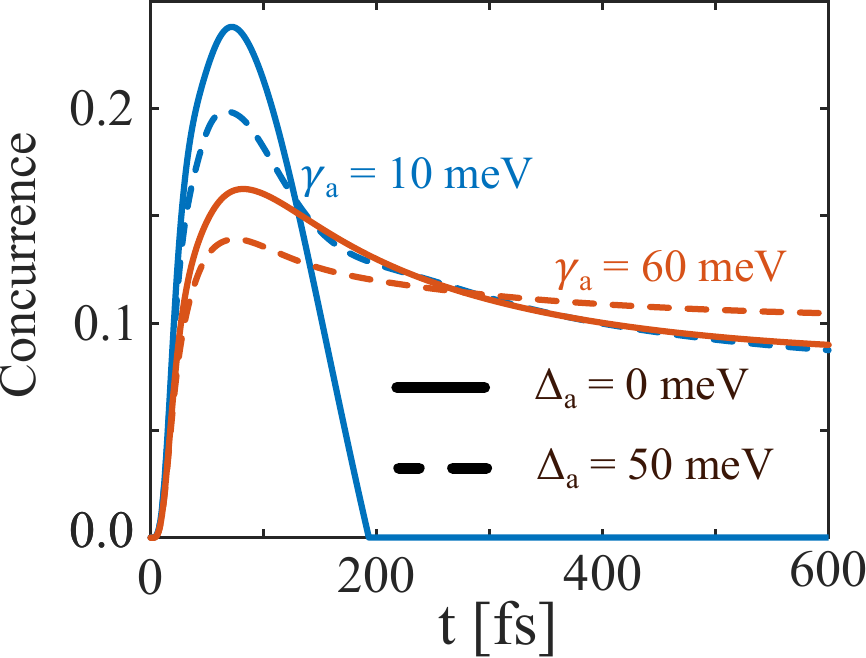}
    \caption{The effect of frequency mismatch between squeezed photons and cavity ($\Delta_a$) on entanglement. Concurrence as a function of time  in presence and absence of cavity-photon frequency mismatch. For small cavity dampings ($\gamma_{a} = 10$ meV) the frequency mismatch facilitates steady-state concurrence. At $\gamma_{a} = 60$ meV frequency mismatch improves the steady-state entanglement.}
    \label{Fig:fig9}
\end{figure}

For parametric-type Hamiltonians it is shown that the range of driving amplitude ($\epsilon$) for which stable steady-state solutions are acquired increases with the increase of cavity-squeeze photons frequency mismatch ($\Delta_a$) and the damping of the constituents~\cite{leroux2018enhancing,zhang2021parity,gardiner2004quantum}. A specific range of $\epsilon$ which produces stable results requires a detailed analysis of Eq.~\ref{Eq:7} and is beyond the scope of the paper. Nevertheless, throughout the paper we have assumed strong damping for plasmons ($Q_{pl} \approx 13$) and photons ($Q_{ph} \leq 200$) to ensure stability. In Fig.~\ref{Fig:fig9} we explore the effect of the other stability parameter ($\Delta_a$) on entanglement. For $\gamma_{a} = 10$ meV a large value of frequency mismatch ($\Delta_a =50$ meV) slightly lowers peak concurrence while it leads to generation of steady-state entanglement. At $\gamma_{a} = 60$ meV frequency mismatch marginally improves steady-state entanglement measured as concurrence. Therefore, the main results of the paper remain valid even in the presence of significant mismatch.
\section{Conclusions}
We demonstrated a nanoparticle-in-cavity propose a scheme to create entanglement between pairs of quantum dots with a squeezed source of light. The system is composed of a PN that couples to the squeezed photons of the cavity and that, in turn, couples to two quantum dots. Exciting the system with squeezed photons leads to population of the Bell state $B_2$ and depopulation of $B_1$ (Eq. \eqref{bell}) creating entanglement of the two 2-level systems as measured by their quantum concurrence. This population/depopulation arises for two reasons: Incoming squeezed photons are generated in even numbers (pairs) and coherent coupling of $B_1$ and $B_2$  to the photonic excitation are out of phase with respect to each other. We show that concurrence can be increased by switching from a pulsed squeezed photon source to a continuous (pumping) source of squeezed light. Achieving maximum concurrence requires that driving amplitude be tuned to an optimum value that is experimentally controllable. We achieved  steady-state entanglement between the QDs  when the quality factor of the cavity was lowered and/or a frequency mismatch was introduced between squeezed photons and the cavity resonance.  We observe an increase in maximum concurrence with the addition of a second PN. Unlike similar schemes that were previously introduced \cite{otten2015entanglement,otten2016origins,lee2013robust,he2012strong, gonzalez2011entanglement,martin2011dissipation,gonzalez2013non}, generation of entanglement in our case does not rely on fine-tuning of QD-plasmon mismatch ($\Delta g_{bc}$). As a result, there is no need for nanoscale precision in positioning the QDs with respect to the dissipating environment.

 Many of the necessary ingredients for realization of our scheme have been realized in separate experiments although, not together in a single study. The photon-plasmon and plasmon-QD couplings used here are less than their documented values at optical frequencies~\cite{ameling2013microcavity,santhosh2016vacuum,yoshie2004vacuum} making experimental implementation more feasible. Recently, photonic cavity based degenerate optical parametric oscillators have been realized~\cite{marty2021photonic}. Our scheme also requires PNs to interact with the squeezed light, e.g few photon absorption and re-emission. This has been suggested in theory~\cite{chang2006quantum,tame2008single,thakkar2015quantum,nascimento2015modeling} and verified in experiment~\cite{altewischer2002plasmon,fasel2005energy,di2012quantum}. Also, the optical response of plasmonic-QD systems is known to be sensitive to the quantum state of the QDs, suggesting entanglement detection via optical cross section measurement~\cite{shah2013ultrafast}. 

To our knowledge, this is the first scheme of entanglement generation by selective damping of Bell states where a quantum source of light is used. However, it should be noted that here the dissipating element (in this case the PN) play other roles in addition to selective damping of Bell states. The PN absorbs the quantum light and facilitates population/depopulation of Bell states $B_2$/$B_1$ in addition to strong selective damping of states $B_{3,4}$. Also, the heavy damping of the PN provides a certain degree of entanglement insensitively to changes in photon-plasmon coupling strength(see Fig.~\ref{Fig:fig6}(c)). Another point of distinction to previous studies is the counter-intuitive role of cavity damping. Cavity damping selectively dissipates certain Bell states so a cavity with low quality factor enables steady-state entanglement of the QDs.

The results presented here are intriguing as such robust entanglement generation in open quantum systems is a challenge~\cite{verstraete2009quantum,kraus2008preparation}. In addition, we achieved steady-state entanglement in plasmonically coupled QDs, but in contrast to other work ~\cite{otten2016origins,cortes2020non}, with finite (i.e nonzero) dephasing rates. Nevertheless, we acknowledge some unresolved issues. For example, the theoretical upper bound limit of concurrence is still unknown. Also, whether or not 
 a time-dependent pattern of squeezed pulse can improve entanglement requires further exploration. The reason behind the concurrence disappearance in the strong photon-plasmon coupling regime and the possible utility of non-degenerate squeezed light for entanglement generation are other questions which demand further investigations. Therefore, the presented work should be viewed as propitious start that encourages future investigations.






\section*{Appendix A: CONCURRENCE}

Throughout the paper  we use an entanglement monotone called concurrence to quantify entanglement between the two quantum dots. Concurrence is an algebraic function of entanglement entropy and is specifically used to benchmark entanglement of bipartite two-level systems~\cite{Wootters1998}. For maximally entangled systems concurrence takes the value of 1 and for systems in separable states concurrence is zero. For density matrix results, concurrence at time $t$ is calculated from the reduced density matrix of the two QDs ($\rho_{c}(t)$). $\rho_{c}(t)$ is calculated by tracing out the plasmonic and photonic degrees of freedom from the density matrix:
\begin{equation}
    \braket{B_i|\rho_{c}(t)|B_j} = \sum_{n_{a}}\sum_{n_{b}} \braket{n_{a},n_{b},B_{i}|\rho(t)|n_{a},n_{b},B_{j}}.
\end{equation}
Concurrence is calculated as
\begin{equation}
Concurrence  = max (\lambda_1 - \lambda_2 - \lambda_3 - \lambda_4,0),
\end{equation}
where $\lambda_{1-4}$ are eigenvalues of the matrix $\sqrt{\sqrt{ \rho_c} \tilde \rho_c \sqrt{\rho_c}}$ in decreasing order. $\tilde \rho_c$ is defined as: 
\begin{equation}
\tilde \rho_c = \left(\sigma_y \otimes \sigma_y\right)\rho_c^*\left(\sigma_y \otimes \sigma_y\right).
\end{equation}
For the pure states resulting from the non-Hermitian Hamiltonian formalism concurrence is equal to  
\begin{equation}
Concurrence = \sum_{i=1}^4 |\beta_i^2|,
\end{equation}
where $\beta_i$ are the probability amplitudes of Bell states in the magic basis 
\begin{subequations}
\begin{equation}
\ket{e_1} = \frac{1}{\sqrt {2}}\left(\ket{0,0} + \ket{1,1}\right) 
\end{equation} 
\begin{equation}
\ket{e_2} = \frac{i}{\sqrt {2}}\left(\ket{0,0} - \ket{1,1}\right) 
\end{equation}
\begin{equation}
\ket{e_3} = \frac{i}{\sqrt {2}}\left(\ket{0,1} + \ket{1,0}\right) 
\end{equation}
\begin{equation}
\ket{e_4} = \frac{1}{\sqrt {2}}\left(\ket{0,1} - \ket{1,0}\right).
\end{equation}
\label{bell2}
\end{subequations}

\section*{Appendix B: Basis for the Schr{\"o}dinger Equation solution}
As explained in the main text only kets with total excitation number of 2 ($ n_{tot} = n_{a} + n_{b} + n_{c1} + n_{c2} = 2$) couple to the initial two-photon pulse. Also, the Hamiltonian in Eq.~\ref{Eq:6} only couples states with the same total excitation number ($\braket{\psi_{n_{tot}}|H_{int}|\psi_{m_{tot}}}=0, m_{tot}\neq n_{tot}$). Therefore, the ket describing the system can be written as a superposition of states with excitation number of 2.
 The ket describing the system can be expanded as: 
\begin{equation}
\begin{split}
\ket{\psi(t)} & = 
\alpha_1'(t)\ket{0,0,1,1} + 
\alpha_2'(t)\ket{0,1,0,1} + 
\alpha_3'(t)\ket{0,1,1,0}  \\
 & +  \alpha_4'(t)\ket{1,0,1,0} +
\alpha_5'(t)\ket{1,0,0,1}  +
\alpha_6'(t)\ket{1,1,0,0} \\
& + \alpha_7'(t)\ket{0,2,0,0} + \alpha_8'(t)\ket{2,0,0,0}. 
\label{Eq:11}
\end{split}
\end{equation}
Since the plasmonic damping is strong, we can assume the two-plasmon state has negligible probability and 
 $\ket{0,2,0,0}$ can be omitted. Using,
\begin{subequations}
\begin{equation}
\ket{n_{a},n_{b},1,1} = \frac{1}{\sqrt 2}\left( \ket{n_{a},n_{b},B_1} - \ket{n_{a},n_{b},B_2}\right)
\end{equation} 
\begin{equation}
\ket{n_{a},n_{b},0,1} = \frac{1}{\sqrt 2}\left( \ket{n_{a},n_{b},B_3} - \ket{n_{a},n_{b},B_4}\right)
\end{equation}
\begin{equation}
\ket{n_{a},n_{b},1,0} = \frac{1}{\sqrt 2}\left( \ket{n_{a},n_{b},B_3} + \ket{n_{a},n_{b},B_4}\right),
\end{equation}
\label{}
\end{subequations}

 \noindent to rewrite the first five kets in  Eq.~\ref{Eq:11}, gives $\ket \psi$ in the custom basis:
 \begin{equation}
\begin{split}
\ket{\psi(t)} & = 
\alpha_1''(t)\ket{0,0,B_1} + 
\alpha_2''(t)\ket{0,0,B_2} + 
\alpha_3''(t)\ket{0,1,B_3}  \\
 & +  \alpha_4''(t)\ket{0,1,B_4} +
\alpha_5''(t)\ket{1,0,B_3}  +
\alpha_6''(t)\ket{1,0,B_4} \\
& + \alpha_7''(t)\ket{2,0,0,0}. 
\label{Eq:12}
\end{split}
\end{equation}
When there is no plasmon-QD coupling mismatch $\Delta g_{bc} = 0$ photonic and plasmonic antisymmetric states do not couple to the rest of the states in Eq.~\ref{Eq:12}:

\begin{subequations}
\begin{equation}
\braket{1,0,B_4|H_{int}|\psi_{n=2}} = 0
\end{equation} 
\begin{equation}
 \braket{0,1,B_4|H_{int}|\psi_{n=2}} = 0.
\end{equation}
\label{bell2}
\end{subequations}
Hence, in the absence of plasmon-QD coupling mismatch the ket describing the system can be expanded as: 
\begin{equation}
\begin{split}
\ket{\psi(t)} & = 
\alpha_1(t)\ket{0,0,B_1} + 
\alpha_2(t)\ket{0,0,B_2} + 
\alpha_3(t)\ket{0,1,B_3}  \\
 & +  \alpha_4(t)\ket{1,0,B_3} +
\alpha_5(t)\ket{1,1,0,0} +
\alpha_6(t)\ket{2,0,0,0}.
\label{}
\end{split}
\end{equation}.

\section*{Appendix C: Hamiltonian and the Lindblad superoperator  for the two-PN two-QD system}

With the addition of a second PN the Hamiltonian changes to  
\begin{equation}
\begin{split}
&  H = \Delta_a \hat{a}^\dagger a + \epsilon(\hat{a}^\dagger \hat a^\dagger + \hat a\hat a)  + \sum_i^2   \Delta_{b}\hat{b_i}^\dagger \hat b_i + g_{ab}^i(\hat a^\dagger \hat b_i+\hat b_i^\dagger \hat a)   \\ &  +\sum_{j=1}^2 \Delta_{c}\hat{c_j}^\dagger \hat c_j +  g_{ac}^i(\hat a^\dagger \hat c_j + \hat c_j^\dagger \hat a) + 
g_{bc}^{ij}(\hat b_i^\dagger \hat c_j + \hat c_j^\dagger \hat b_i).
 \end{split}
 \end{equation}
 Also, the Lindblad superoperator acquires an extra plasmonic damping term and becomes
 \begin{equation}
     \hat L(\hat \rho)  = \hat L_{\hat a}(\hat \rho)  + \hat L_{\hat b_1}(\hat \rho) + \hat L_{\hat b_2}(\hat \rho) + \hat L_{\hat c_1^+ \hat c_1}(\hat \rho) + L_{\hat c_2^+ \hat c_1}(\hat \rho).
 \end{equation}
 We assume the two PNs are identical which means the PNs damping rates are equal. Also, we assume they couple with the the same strength to the quantum dots and photons ($g_{ab}^1 = g_{ab}^2$ and $g_{bc}^{1j} = g_{bc}^{2j}$). Plasmon-plasmon coupling is neglected.

\bibliography{apssamp}

\end{document}